# Entrance Channel X-HF (X=Cl, Br, and I) Complexes studied by High-Resolution Infrared Laser Spectroscopy in Helium Nanodroplets

Jeremy M. Merritt, Jochen Küpper[*] and Roger E. Miller[†]

Department of Chemistry

University of North Carolina,

Chapel Hill, NC 27599, USA

Rotationally resolved infrared spectra are reported for halogen atom – HF free radical complexes formed in helium nanodroplets. An effusive pyrolysis source is used to dope helium droplets with Cl, Br and I atoms, formed by thermal dissociation of $Cl_2$, $Br_2$ and $I_2$. A single hydrogen fluoride molecule is then added to the droplets, resulting in the formation of the X-HF complexes of interest. Analysis of the resulting spectra confirms that the observed species have $^2\Pi_{3/2}$ ground electronic states, consistent with the linear hydrogen bound structures predicted from theory. Stark spectra are also reported for these species, from which the permanent electric dipole moments are determined.

[*] Present address: Fritz-Haber-Institut der MPG, Faradayweg 4-6, D-14195 Berlin, Germany

[†] Electronic mail: remiller@unc.edu

**Introduction**

The effects of the moderately strong long-range electrostatic interactions in ion-molecule reactions are well known to have a significant influence on the associated reaction rates. In contrast, the importance of the weaker van der Waals forces in the entrance channels of neutral reactions have been only recently fully appreciated.[1-4] For example, experimental and theoretical work on the Cl + HD reaction[5-8] shows that the torque experienced by the HD in the entrance valley of the potential has a significant effect on the overall reaction rates and branching ratios. Thus, two potentials that are rather similar in the transition state region display quite different reaction dynamics, owing to the fact that the one is repulsive at long range (particularly in the perpendicular geometry), while the other is attractive (0.5 kcal/mol), with a T-shaped equilibrium geometry. Studies of this type emphasize the importance of characterizing the pre- and post- reactive regions of these potential energy surfaces.

In recent years a number of methods have been developed for probing pre-reactive complexes using spectroscopic methods.[1,9,10] The fast cooling associated with a free jet expansion can stabilize these complexes, even though the barriers to reaction may be quite low. Spectra of these open shell complexes provide detailed information on the associated potential energy surfaces, using methodologies quite similar to those developed for the study of closed shell van der Waals complexes.[11] There is also considerable promise for the study of the dynamics of these pre-reactive complexes, namely regarding the photo-initiation of the associated chemical reactions. In particular, the barriers to reaction are often sufficiently low so that vibrational excitation of the complex has the potential to initiate the reaction.[1,12,13]

An important example of this is HO-CO,[1] a precursor complex for the well-studied CO + OH $\rightarrow$ $CO_2$ + H reaction, of fundamental importance in combustion and atmospheric processes. By probing the intermolecular vibrations of this complex, it has been possible to map out much of the pre-reactive region of the potential energy surface. Intramolecular vibrational excitation, of the O-H stretch for example, can lead to either reaction, to form the H + $CO_2$ products, or vibrational predissociation into OH + CO. To date, the experimental results are inconclusive as to which channel is most important. Nevertheless, the potential of these methods for studying photoinduced chemical reaction dynamics is great.

Matrix isolation methods have also been used to stabilize reactive species.[14-16] Classical cryogenic noble gas matrices have several advantages, including controlled doping and accumulation times. Large numbers of the radicals can often be obtained due to isolation of the reactants in the inert cage of the matrix, which limits their reactivity. Annealing the matrix provides some mobility, so that experiments can be carried out before and after the reaction takes place. The primary difficulty here is that the associated spectroscopy has rather poor resolution due to the matrix effects, thus limiting the information that can be extracted. In recent years, helium nanodroplets have been shown to provide many of the advantages of conventional matrices, with the added benefit of high spectral resolution.[17] Indeed, rotationally resolved vibrational spectra can be obtained for complexes solvated in helium.[18] These complexes are easily grown in the helium nanodroplets by simple pick-up of two or more molecules or atoms. The high mobility of the molecules within the droplets and the low temperature (0.4 K) ensures that they quickly form a complex. As a result of the rapid cooling provided by the helium, the systems of interest are often trapped in local minima on the potential energy surface.[19,20] In the case of reactive systems, one expects that this rapid cooling will also allow trapping of pre-reactive species, with very shallow wells in the entrance channel to the associated chemical reaction. In this work we explore the possibility of using helium nanodroplets to study radical complexes, with specific application to the X-HF (X=Cl, Br and I) complexes formed by independent pick-up of the corresponding atoms and molecules by the nanodroplets.

Heavy-light-heavy systems (X-HY, X & Y = F, Cl, Br and I) figure prominently in reaction dynamics studies, being prototypical systems for the observation of transition state resonances. Theoretical studies on these X-HY systems[21-28] have revealed many important issues related to the role of multiple electronic surfaces and spin-orbit interactions. These systems have also been the subject of considerable experimental study, including the spectroscopy of the transition states, accessed by photodetaching an electron from the [XHY]$^-$ anion complexes,[29] and bond-specific photo-dissociation of HX dimers.[10,30] In addition, evidence for the existence of weakly bound X-HY complexes has come from infrared spectroscopy in noble gas matrices.[15,31-33] Unfortunately, these studies do not provide detailed information on the associated entrance channel wells.

As illustrated by the Cl + HD system discussed above, the structures of entrance channel complexes (which probe the associated potential energy surfaces at energies that are relevant to the reactive dynamics) are of considerable importance in understanding the nature of the reactions, particularly at low translational energies. The need for new experimental methods for probing such species is clear, given the lack of data on these X-HY systems, despite the intense interest in the associated reactions. On the one hand, the interaction between the large dipole moment of the HY molecule and the quadrupole moment of the halogen atom (X) would suggest two linear structures, namely a hydrogen bonded geometry and one with the two heavy atoms adjacent to one another. On the other hand, the HY molecules have substantial quadrupole moments, so that T-shaped structures are also predicted to be minima. It is actually quite surprising that the structures of these systems have not yet been conclusively determined, illustrating the primitive state of our understanding.

High-level *ab initio* and semi-empirical calculations have been performed for several of these systems.[21-28] The weak interactions between a halogen atom and a HY molecule suggest that, to a first approximation, the orbital angular momentum of the single atom (l) will be conserved. In this case there are three diabatic surfaces that correlate with the unpaired electron being in the $p_x$, $p_y$, or $p_z$ orbital. Taking the z direction as the molecular axis, two of these configurations are degenerate for linear configurations, but diverge as the molecule bends (a Renner-Teller effect). Recent calculations by Zeimen *et al.*[28] predict that the global minimum for the Cl – HCl system is T-shaped, in contrast with a linear prediction made in earlier studies,[21,34] once again illustrating the need for experimental measurements of the associated structures.

High-resolution spectroscopy is often looked upon as the definitive tool to determine accurate molecular geometries. While this statement is well supported by studies of a wide range of cluster systems, the situation is less clear for the X-HY complexes. Bound state calculations on Cl-HCl show distinct progressions of both linear and T-shaped states, even though both are believed to display the spectral signatures of a linear open-shell molecule.[28] This peculiarity is quite unique, suggesting that in some cases the high-resolution spectroscopic experiments which provide rotational constants, dipole moments, parity splitting, etc … may not give conclusive structural information. Elucidating the structures and energetics of these complexes will clearly require extensive interactions between experiment and theory.

**Experimental**

The helium droplet apparatus used in the present study is shown schematically in Figure 1. For a more detailed description the reader is referred to reference 35. Helium gas is expanded from a 5 μm orifice that is cooled to 19–24 K and maintained at a stagnation pressure of 50 – 90 bar, resulting in the production of helium nanodroplets with a mean size of typically 3000 - 7000 atoms. The species of interest are added to the nanodroplets via the pick-up method,[35] in this case in two separate regions. A simple low-pressure pick-up chamber is used to add the HF molecules, while the radicals are produced via pyrolysis, as discussed below.

A rich literature already exists that deals with the study of radicals produced by pyrolysis,[36-38] microwave discharge,[39-44] electric discharge,[45-48] and photolysis.[49-56] Although these previous approaches provided us with considerable guidance in the development of a radical source for helium nanodroplet doping, none were directly applicable to the current application, which requires low pressures for pick-up ($<10^{-4}$ mbar). The source used here is similar to the one reported previously by our group for the spectroscopic study of the propargyl radical.[57] An effusive source is appropriate for helium nanodroplet studies given that the latter provide the needed cooling, making the free jet expansion of the radicals unnecessary.

The pyrolysis source consists of an open-ended double bore alumina tube (99.8% purity $Al_2O_3$) with two 1mm inner diameter holes oriented along the droplet beam axis. The end section (~1cm) is wrapped with a 0.25mm diameter tantalum filament and encased in ceramic paste. Typically, 30 Volts (5 Amps) is sufficient to reach temperatures of ~ 1600K, measured by a K-type thermocouple. At this temperature the source can be routinely operated for periods in excess of one hundred hours. The exit of the source is positioned just downstream of the skimmer used to collimate the helium nanodroplet beam. The nanodroplets pass within 1 mm of the exit of the pyrolysis source. The helium droplets are virtually transparent to the black body radiation from the source, so that there is no discernable evaporation of the droplets, even when the source is heated to 1600 K. At these temperatures the halogen molecules are dissociated into atoms with nearly unity efficiency. The dissociation process was monitored by spectroscopically probing the $X_2$-HF and X-HF species, as a function of the source temperature. The second pickup region was located downstream of the pyrolysis source and was used to add a single HF molecule to the droplets. The high mobility of the molecule and halogen atom within the nanodroplet ensures that they quickly find one another to make the complexes of interest, the associated condensation energy being quickly dissipated to the surrounding helium, resulting in the evaporation of a small fraction of the helium atoms from the nanodroplets.

The doped nanodroplets enter the laser excitation region somewhat further downstream. A tunable infrared F-center laser (Burleigh FCL-20) is reflected between two parallel gold-coated mirrors to increase the length of the laser-helium nanodroplet interaction region. In the present study, the laser was operated on crystal #2 (KCl:Li) and was pumped by a krypton ion laser. Vibrational excitation of the H-F stretch in the X-HF complex is followed by vibrational relaxation to the surrounding helium, resulting in the evaporation of several hundred helium atoms. A bolometer detector is used to measure the corresponding reduction in the helium droplet beam flux. The laser beam is amplitude modulated and the signals are processed by a phase-sensitive detector.

A large electric field can be applied to the laser interaction region in order to collapse the rotational fine structure in the infrared spectrum of the complex. This pendular state method[58,59] provides a considerable improvement in the sensitivity of the experiment and was used when originally searching for new species. At somewhat lower electric fields, Stark spectroscopy can also be performed to determine the permanent electric dipole moment of these complexes.

**Electronic Structure Calculations**

As noted above, there have been numerous theoretical studies of X-HY systems, although the majority of these have been focused on the symmetric X-HX systems.[21-24,26-28] However, Meuwly and Hutson recently reported semi-empirical calculations for the Br-HF system.[60] Adiabatic potential surfaces including spin-orbit coupling predict the global minimum for this system to be linear (-670 cm$^{-1}$), with an intermolecular hydrogen bond. The ground electronic state of this system is predicted to be $^2\Pi_{3/2}$. A T-shaped minimum (395 cm$^{-1}$ higher in energy) is also predicted, although the barrier connecting this well to the linear one is quite small ( <50 cm$^{-1}$). Recent calculations on Cl-HCl[28] in contrast has a T-shaped global minimum. The difference between these two cases is attributed to the large dipole of HF (compared to that of HCl) and the large polarizability of the Bromine atom (compared to that of Chlorine). As a result, the linear minimum is considerably more stable in the Br-HF case. To date, however, there have been no experimental confirmations of these predictions.

Although studies near the transition state region of the potential surface often require multireference calculations, to include the effects of the low lying excited electronic states, the present study of the entrance channel region of the potential, where the ground state dominates, single reference calculations are expected to be sufficient for harmonic vibrational frequencies and structural parameters (dipole moments, rotational constants, etc...). *Ab initio* calculations were performed here using Gaussian 03[61] at the UMP2 and UCCSD(T) level incorporating the Aug-cc-pVTZ basis of Dunning.[62] The results of geometry optimizations and harmonic frequency calculations are summarized in Tables 1-3 for F-HF, Cl-HF and Br-HF. Binding energies were further corrected for basis-set superposition error using the method of Boys and Bernardi.[63] In agreement with previous studies, we find three stable minima corresponding to the two linear and T-shaped complexes. The calculated global minima for these systems all correspond to linear hydrogen bonded structures. Comparison of the present binding energies, with those reported previously for Br-HF,[60] reveals that the current potential has a somewhat deeper T-shaped minimum (-404 cm$^{-1}$ vs. -275 cm$^{-1}$), while the linear minimum (hydrogen bonded) are quite similar (-720 cm$^{-1}$ vs. -670 cm$^{-1}$). Calculations for I-HF were also carried out using a correlation consistent basis set, in conjunction with a small core relativistic pseudopotential (aug-cc-pVTZ-PP).[64] In this calculation, 28 core electrons of iodine were replaced by an effective core potential, while the remaining bonding electrons and the HF molecule were treated within the aug-cc-pVTZ basis.[62] Similar calculations were carried out for bromine (10 frozen core electrons) for comparison with the all electron calculations.

Calculations were also performed for the Br$_2$-HF system, to aid in identifying the precursor complexes, which in turn helped in optimizing the pyrolysis source. Figure 2 shows a two-dimensional cut through the associated coupled rotor potential, calculated at the MP2 / aug-cc-pVDZ level of theory. For each point, the two rotor angles, theta and phi were stepped incrementally (10°), while all other geometric parameters were relaxed. The energies were corrected for basis set superposition error (BSSE) using counterpoise correction[63] and a bi-cubic spline interpolation was used to generate the surface. In agreement with the experimental results discussed below, we find two minima on the potential energy surface, one corresponding to a L-shaped, hydrogen bonded complex and the second to a bent isomer bonding at the fluorine end of the HF molecule. The results from fully relaxed optimizations, as well as harmonic frequency calculations, for these two isomers are summarized in Table 4. Comparisons with the experimental results are given below.

**Experimental Results**

**The $Br_2$ - HF Complexes**

The process of optimizing the experimental conditions began with the search for the $Br_2$-HF complexes discussed above, yielding the two spectra shown in Figure 3. Both of these spectra optimized under the conditions needed for the pick-up of a single molecule of each type (namely HF and $Br_2$), confirming that they are associated with the 1:1 complex. The significant frequency shift (-85 $cm^{-1}$) for the band in Figure 3A (relative to the HF monomer at 3959.19 $cm^{-1}$ [65]) suggests that this is the hydrogen bonded complex. Indeed, the calculated asymmetric top spectrum, shown in Figure 3B, is consistent with the L-shaped isomer discussed above, with the rotational constants reduced to account for the effects of the helium.[17] Due to the small asymmetry associated with this complex the fit to the spectrum provides only $(B+C)/2$, as given in Table 4. The experimental frequency shift for this isomer is also in reasonable agreement with the *ab initio* calculations (-100 $cm^{-1}$), where the latter were first scaled to give the correct vibrational origin for HF monomer in helium.

The spectrum in Figure 3C has a much smaller frequency shift (-17.72 $cm^{-1}$, relative to the HF monomer) in excellent agreement with the calculated frequency shift (scaled to HF monomer) for the bent isomer, namely -20.43 $cm^{-1}$. This spectrum is well represented by a linear rotor calculation (Figure 3D), consistent with the fact that the bent isomer has a very large A rotational constant. As a result, only the $K_a = 0$ levels are populated at the temperature of the droplets (0.37 K). The resulting $K_a = 0 \leftarrow 0$ band is indistinguishable from that of a linear molecule spectrum and again we report only $(B+C)/2$. The rotational constants used to fit this spectrum are approximately a factor of five smaller than the *ab initio* calculations, due to the effects of the helium solvent. Experimental parameters derived from the simulations, along with *ab initio* data are presented in Table 4 for both isomers.

**The Br - HF Complex**

Having identified the spectra associated with the $Br_2$-HF 1:1 complexes, we can use these as an aid in optimizing the conditions for the production of the Br-HF complex. Figure 4B shows a pendular spectrum of the bonded H-F stretching region, recorded with the pyrolysis source at room temperature. The peak near 3862 $cm^{-1}$ is well known from our previous study[66] to be due to the hydrogen bonded H-F stretch of the HF dimer, while the peak at the high frequency end of the scan is due to the L-shaped $Br_2$-HF complex discussed above. In both cases, the electric field has collapsed the rotational structure into a single peak. Figure 4A shows a scan of the same region of the spectrum, recorded with the pyrolysis source heated to 1200 K.

Note that the pendular transition associated with the Br$_2$-HF complex is diminished in intensity, relative to that obtained with a room temperature source, and that a new peak has appeared in the spectrum. The growth of this band is complementary to the disappearance of the spectra assigned to both of the Br$_2$-HF isomers. This new band optimizes under conditions that correspond to, on average, the capture of a single bromine atom.

Having located a vibrational band that is tentatively assigned to the Br-HF complex, the rotational structure in the band can be examined by turning off the electric field. Figure 5A shows the corresponding zero field spectrum, which has clearly resolved P, Q and R branches. There is a substantial "gap" between the Q branch and the corresponding P and R branches, consistent with this being a linear complex with a Π ground electronic state. Indeed, the $^2\Pi_{3/2}$ state is expected to give rise to a Q branch separated from the nearest P and R branch transitions by ~5B. This is in turn consistent with the fact that the ground electronic state of the Br atom is $^2P_{3/2}$.

Adopting a $^2\Pi_{3/2}$ model, we can simulate the spectra using the corresponding Hamiltonian. In particular, the rotational energy levels are given by $E(J) = B[J(J+1) - \Omega^2] - D[J(J+1) - \Omega^2]^2$, where B and D are the rotational and centrifugal distortion constants, respectively, and Ω is the quantum number associated with projection of the total electronic angular momentum, which for the ground state is $^3/_2$. The total angular momentum quantum number (J) must be greater than or equal to Ω ($^3/_2$). As a result, the J = $^1/_2$ rotational level is not allowed, accounting for the missing R($^1/_2$) and P($^3/_2$) transitions, which would otherwise appear in the gap between the Q branch and the R and P branches, respectively. The spectroscopic constants (ν, B, and D) resulting from the fit to the spectrum (see Figure 5B), are summarized in Table 5. The rotational temperature was determined by fitting the relative transition intensities, yielding 0.35 K, in good agreement with previous results on closed shell systems.[67] The experimental (helium) rotational constant is a factor of 2.2 smaller than the *ab initio* value.[68] This ratio is slightly low, but certainly consistent with what has been observed for many other systems.[69] The experimental monomer-to-complex frequency shift is determined to be -91.73 cm$^{-1}$, which is in good agreement with the (scaled) all electron *ab initio* (CCSD(T)) calculation for the linear hydrogen bound Br-HF isomer, namely -103.2 cm$^{-1}$. In comparing these values we assume that the effects of the helium on the vibrational frequency is small. For such a hydrogen bound species we indeed expect that the vibrational origins will be shifted by less than 10 cm$^{-1}$ from the gas phase values.[70] The present work also supports the assignment made previously in argon matrices (-128 cm$^{-1}$),[15] noting the significantly larger matrix shift in this case.

Although the simple model presented above provides a reasonable qualitative description of the experimental spectrum, there is clearly fine structure in the experimental spectrum that is not reproduced by the simulation. It is important to point out that unusual lineshapes and even splitting of transitions can result from the interactions between the molecules of interest and the helium solvent.[35,71] However, these effects are usually highly dependent upon the helium nanodroplet size, which is not the case for the Br-HF spectrum considered here. For this reason, we explored other possible sources for the observed fine structure, which is most evident in the transitions associated with low J states. We also considered the fact that bromine has two isotopes of significant natural abundance, namely $^{79}$Br (50.69%) and $^{81}$Br (49.31%). Nevertheless, the corresponding moments of inertia are not different enough to explain all of the structure observed in the experimental spectrum. In contrast, the isotope effects for the lighter Cl-HF complex (discussed below) are quite significant.

For closed shell molecules, where the magnetic fields at the nuclei are zero, the leading terms in the nuclear hyperfine structure are often those associated with nuclear electric quadrupole effects. Some effort was made to fit the Br-HF spectrum with the corresponding Hamiltonian, with only limited success. For this open shell system it is more likely that the dominant hyperfine structure results from the interaction of the magnetic field associated with the unpaired electron with the nuclear magnetic dipole moment. Indeed, the two isotopes of bromine have large and comparable nuclear magnetic dipole moments.[72]

The angular momentum coupling cases for complexes containing open-shell atoms have been described previously by Dubernet and Hutson.[73] In their notation, the X-HF complexes belong to case one, which assumes that, upon complexation, the large spin-orbit interaction of the halogen atom is not significantly changed. The differences between Dubernet and Hutson's case one, and Hund's case A for diatomic molecules, however, would be indistinguishable in our experiment, so we resort to a treatment of the nuclear hyperfine interaction for Hund's case A. Following the treatment of Frosch and Foley,[74] the Hamiltonian describing the interactions pertinent to Hund's case A, can be written as: $H' = a\Lambda(I \cdot k) + b(I \cdot S) + c(I \cdot k)(S \cdot k)$, where k is a unit vector along the molecular axis. Assuming $\Lambda$ and $\Sigma$ are good quantum numbers associated with the projections of the orbital and spin angular momenta, respectively (the definition of Hund's case A), this expression can be reduced to $H' = [a\Lambda + (b+c)\Sigma]$ I·k.[72] The interaction energy is then: $W = [a\Lambda + (b+c)\Sigma]\frac{\Omega}{J(J+1)} I \cdot J$, where $I \cdot J = \frac{F(F+1) - J(J+1) - I(I+1)}{2}$. The relevant selection rules for this case are $\Delta F = 0, \pm 1$ and $\Delta J = 0, \pm 1$, with F' = 0 ←/→ F" = 0. The quantum number associated with the nuclear spin angular momenta (I) is $3/2$. (Note that $I = 3/2$ for chlorine as well, while for iodine, $I = 5/2$). J is the quantum number associated with the total angular momentum exclusive of nuclear spin, and F is the coupling of J and I.

The simulated spectrum for Br-HF, including nuclear magnetic hyperfine coupling, is shown in Figure 6B, compared to the experimental spectrum in Figure 6A. A stick spectrum is also shown below the simulated spectrum to illustrate the makeup of the various features in the spectrum. The addition of these hyperfine terms in the Hamiltonian clearly give results that reproduce all of the important features in the experimental spectrum. The effective hyperfine constant $[a\Lambda + (b+c)\Sigma]$ used in the simulation is given in Table 5, namely 0.045 cm$^{-1}$.

The nuclear magnetic hyperfine splitting can be used to study the unpaired electron density at the bromine nucleus. A comparison of these effects for a van der Waals complex and the isolated atom is difficult, however, and very few examples of this exist in the literature. Important exceptions are the Ar-OH[75] and Ar-OD[76] complexes. In these cases the ground electronic state (X $^2\Pi$) hyperfine parameters are found to be virtually unchanged from the free molecules. In contrast, in the A $^2\Sigma^+$ electronically excited state of OH, a small change in the nuclear hyperfine interaction was used to argue that there was a slight change in the chemical bonding. Unfortunately, the lack of experimental data for bromine atom containing van der Waals complexes precludes any quantitative comparisons for the Br-HF system. Clearly this is an area that deserves more in-depth theoretical study.

The intermolecular bending and stretching levels of Br-HF were recently calculated,[60] leading us to search for the corresponding combination bands with the H-F stretch. These states, which correspond to excitation of intermolecular bending and stretching modes are particularly interesting given that they are sensitive to all 3 diabatic surfaces ($1^2A'$, $2^2A'$, and $1^2A''$).[60] The combination bands are predicted to be 145 and 172 cm$^{-1}$ above the H-F fundamental (3867.4 cm$^{-1}$), for the $|P| = ½$ and $^5/_2$ states, respectively. Unfortunately, extensive searches in the region between 4000 – 4050 cm$^{-1}$ did not reveal these bands. This may be due to the fact that the F-center laser has relatively little power in this region. Further work in this direction is planned using a higher power PPLN-OPO laser system.[77]

**The I – HF Complex**

A similar study was carried out for the I-HF complex, for which $^{127}$I is the only naturally occurring isotope. Figure 7 shows a comparison between the experimental and simulated spectra, including the hyperfine effects discussed above. The molecular constants obtained from this fit are given in Table 5. Here again, the experimental rotational constants are consistent with those from the *ab initio* calculations, corresponding to a ratio of ~2.2. For this case the experimental monomer-to-complex frequency shift is determined to be -113.6 cm$^{-1}$, which is in reasonable agreement with the calculations using the effective core potentials (-150 cm$^{-1}$). Qualitatively, the features in this spectrum are comparable with those for the Br-HF complex, again indicating that the vibrationally averaged structure is linear. The nuclear magnetic hyperfine constants derived for the isolated atoms of iodine and bromine are very similar, $^{127}$I = 827.3 MHz,[78] $^{81}$Br = 953.77 MHz, and $^{79}$Br = 884.81 MHz.[79] It is therefore reasonable that the simulation of the Br-HF and I-HF spectra require coupling constants of approximately the same ratio (See Table 5).

**Dipole Moments**

A series of Stark spectra were also recorded for the Br-HF and I-HF complexes. The data shown in Figure 8 for the Br-HF complex was obtained with the laser polarization aligned parallel to the applied DC electric field, yielding $\Delta M_F = 0$ (weak field limit) selection rules. Of particular interest are the two peaks that lie to either side of the main Q branch, which split apart as the field is increased. As shown by the inset in the figure, the splitting between these two peaks increases linearly with field, suggestive of a first order Stark effect that would be expected for a $\Pi$ electronic ground state.[72]

The electric fields used to obtain the spectra in Figure 8 are moderately large, which precludes the use of a perturbation treatment for the Stark effect. Instead, the full Hamiltonian matrix was numerically diagonalized to obtain the corresponding eigenstates and eigenvalues. In the limit of weak electric field coupling, the nuclear spin angular momentum remains coupled to the rotational angular momentum and a coupled basis set ( $|J,\Omega,F>$ ) is appropriate. In this limit, the application of the electric field does not disrupt the coupling of the angular momenta and $M_F$ (the projection of F in the field direction) remains a good quantum number. At higher electric fields, however, the nuclear spin and electron spin become decoupled from the rotation of the molecule. In this limit the molecular eigenstates are best represented in the decoupled representation ($|J,\Omega,M_I>$), where the nuclear and electron spins form a constant projection along the electric field direction, namely $M_I$ and $M_J$, respectively.

Although the use of moderately large electric fields in the present study would suggest the use of the decoupled representation, both basis sets are complete and the eigenvalues obtained with either are the same to within the numerical precision. In fact, the choice of basis set only affects the labeling of states. Since the present spectra are not fully resolved (into individual transitions) the quantum number labels are not that important, so for simplicity we used the coupled representation. The effective Hamiltonian can be written as: $H_{eff} = H_R + H_{CD} + H_{HFS} + H_E$, corresponding to the contributions from rotation, centrifugal distortion, nuclear hyperfine, and the Stark effect, respectively. The corresponding matrix elements are given in the literature.[80] In spherical tensor notation the Stark term can be written as: $H_E = -T^1(\mu_e) \cdot T^1(E)$, which for the present case has the corresponding matrix elements;

$$< J,\Omega,I,F,M_F | -T^1_{p=0}(\mu_e)T^1_{p=0}(E) | J',\Omega,I,F',M_F >$$

$$= -\mu_0 E_0 (-1)^{F-M_F} \begin{pmatrix} F & 1 & F' \\ M_F & 0 & M_F \end{pmatrix} (-1)^{F'+J+I+1} \{(2F'+1)(2F+1)\}^{1/2}$$

$$\times \begin{Bmatrix} J' & F' & I \\ F & J & 1 \end{Bmatrix} (-1)^{J-\Omega} \begin{pmatrix} J & 1 & J' \\ -\Omega & 0 & \Omega \end{pmatrix} \{(2J'+1)(2J+1)\}^{1/2}$$

Calculations were carried out with a coupled basis, $|J,\Omega,F>$ of sufficient size to ensure that the eigenvalues had converged to better than $10^{-3}$ cm$^{-1}$. All of the field free constants were held fixed during the fit to the Stark spectra, so that the only adjustable parameters were the dipole moments in the ground and excited vibrational states. Figure 9 shows a Stark spectrum of the Br-HF complex at an applied electric field of 5.17 kV/cm. The fitted spectrum is in excellent agreement with experiment, yielding a ground state dipole moment of 2.1 ± 0.1 Debye and a change upon vibrational excitation of +0.25 ±0.05 D. In analogy with the Ar-HF complex,[81] this large change in the dipole upon vibrational excitation is most likely due to a increase in the anisotropy of the intermolecular potential, resulting in a decrease in the corresponding bending amplitude. This in turn results in an increase in the vibrationally averaged dipole moment in the excited state.

The fact that the experimental dipole moment is considerably smaller than the *ab initio* value calculated here (2.48 D, corresponding to a rigidly linear complex) lends supporting evidence to large amplitude motions, which act to average the equilibrium linear geometry. A first approximation to this effect on the dipole moments and band origin shifts can be estimated by calculating their expectation values. As shown in Figure 10, at each point along a one-dimensional bending coordinate (represented by rotating the HF about its center of mass, making an angle theta between the A axis of HF and the halogen atom) the resulting energy, dipole moment, and harmonic vibrational frequency were calculated at the UMP2 / Aug-cc-pVTZ level. As the complex bends the degeneracy in the spatial orientation of the unpaired orbital is lifted, resulting in two states, $^2A'$ and $^2A''$ respectively. Based upon simple intuition of electrostatic forces, and confirmed by *ab initio* UMP2 / Aug-cc-pVTZ calculations, the $^2A'$ state (in which the orbital that contains the unpaired electron is in the plane of the bending coordinate) is lower in energy than the $^2A''$ state (except where they are degenerate). The difference potential ($^2A'' - ^2A'$) becomes significant only for large bending angles (176 cm$^{-1}$ at 40 degrees), and thus for simplicity we have only included the $^2A'$ state in our discussion. The bending wavefunctions were calculated numerically using the Numerov-Cooley method,[82] and used to find expectation values given by: $\langle X \rangle = \dfrac{\int_0^\pi \psi X \psi \mathrm{Sin}\theta\, \partial\theta}{\int_0^\pi \psi\psi \mathrm{Sin}\theta\, \partial\theta}$. This treatment yields values of $\langle\mu\rangle$ = 2.40 D and $\langle\nu\rangle$ = 4004 cm$^{-1}$. These values can be compared directly with those obtained from the equilibrium calculation, namely: $\mu_{eq}$ = 2.48 D and $\nu_{eq}$ = 3998 cm$^{-1}$. As one can see, the vibrational averaging lowers the dipole moment and shifts the harmonic vibrational frequency by ~6 cm$^{-1}$ to the blue, in both cases improving the agreement with experiment. Quantitative agreement is not expected, however, given that the single reference *ab initio* calculations do not correctly treat the excited electronic states, which will be important in determining the potential surface away from the minimum.

It is interesting to note that a recent calculation by Meuwly and Hutson[60] gave a dipole moment of 1.21 D for Br-HF, including the effects of vibrational averaging. However, this calculation neglected the induced dipole on the bromine. Given that the dipole moment of the HF monomer has been measured to be 1.819 D,[83] this difference (0.609 D) reflects the effect of vibrational averaging on their potential. If we subtract this value from the equilibrium *ab initio* calculation presented here (which does include the induced dipole on the bromine atom) one predicts a dipole moment smaller than observed (1.87 D vs. 2.1 D), indicating that the anisotropy of their potential surface is too small. In contrast, the vibrational averaging based upon the present *ab initio* surface gives a dipole reduction that is too small (0.08 D), suggesting that the corresponding anisotropy is too large. Future studies should be directed at more accurately determining this contribution so that more quantitative comparisons can be made using a full dimensional potential surface.

It is also important to consider that the influence of the helium solvent on the measurement of the dipole moment has been addressed previously in our group.[84] We found that for a rigid molecule the solvent effect results in a ~1% deviation from the gas phase dipole moment. However, for a more floppy system of the type considered here, the effect of the solvent could be more important. In particular, the presence of the solvent could make the potential surface more anisotropic, giving rise to a larger vibrationally averaged dipole moment. It is interesting to note, however, that experiments from our group on the Ar-HF binary complex in helium droplets[85] show very small helium solvent effects, which we take as evidence that the present results for the X-HF systems are essentially the same as those for the gas phase complexes. This said, the more subtle effects on these observables still need to be examined in detail, to further our understanding of the solute-solvent interactions in this exotic medium.

Similar Stark spectra were recorded for the I-HF system, as shown in Figure 11. The fitted spectrum was calculated with a dipole moment of 2.2 ± 0.1 Debye and a change of +0.2 ±0.05 D upon vibrational excitation. These values are consistent with the trends seen in the *ab initio* calculations.

**The Cl – HF Complex**

Chlorine atoms are more difficult to produce by pyrolysis, requiring that the source be operated at higher temperatures (~1600 K). The spectrum of Cl-HF, obtained under these conditions, is shown in Figure 12. All of the tests discussed above, including the careful examination of the pick-up cell conditions required to optimize the signals, were used to confirm that this is indeed the spectrum of the binary complex. The qualitative features in this spectrum are similar to those observed for the above systems, in particular, the presence of a Q branch and the "gaps" between it and the P and R branches. Although this is consistent with a **$^2\Pi_{3/2}$** ground state, the fine structure observed in the spectrum is quite different from that in the other systems. As illustrated in Figure 12, the entire spectrum appears to be doubled, with each strong transition having a weaker companion on the low frequency side. Indeed, the simulated spectrum in Figure 12, consisting of a sum of two shifted (by 0.0380 cm$^{-1}$) **$^2\Pi_{3/2}$** bands (also shown in Figure 12), accurately reproduces the experimental spectrum. The relative intensities of the two bands are in excellent agreement with the ratio of the natural abundances for $^{35}$Cl and $^{37}$Cl, namely 3.129.[86] Indeed, we fit the central Q branch feature to two Lorentzian functions (see Figure 14), which gave an experimental ratio of the integrated areas of 3.174. Apparently the isotope shift is large enough in this system to observe the corresponding splitting of the transitions. It is interesting to note that although no hyperfine splittings are observed in the spectrum, a nonzero coupling coefficient was necessary to accurately reproduce the rotational line intensities, yielding a nuclear hyperfine coefficient of $[a\Lambda + (b+c)\Sigma] = 0.005$ cm$^{-1}$. Once again, the relative magnitude of this coefficient is in quite good agreement with the relative hyperfine constants for the isolated atoms ($^{35}$Cl=205.29 MHz, $^{37}$Cl=170.69 MHz).[87]

A series of Stark spectra were also recorded for the Cl-HF complex (Figure 13) with similar field strengths as used for Br-HF. The top spectrum of this series was recorded at an electric field strength of ~ 60 kV/cm, corresponding to the pendular regime, so that the P and R branches are no longer visible. In contrast to Br-HF, no discernable field dependent splitting is observed near the Q branch, which supports the assignment of the zero field splitting to an isotope effect, rather than nuclear hyperfine interactions. Figure 14 shows a comparison between an experimental Stark spectrum recorded at 3.5 kV/cm and a simulated spectrum obtained by holding all of the field free parameters fixed, yielding a dipole moment of 1.9 D. The calculated spectrum is a simple addition of the two bands arising from the $^{35}$Cl and $^{37}$Cl isotopomers, the intensities having been scaled in accordance with the chlorine isotope natural abundance ratio. At the spectroscopic resolution available here, the isotopic dependence of the rotational constants and dipole moments could not be determined. In contrast to the bromine and iodine results, the change in dipole moment upon vibrational excitation was found to be less than our experimental accuracy. This result may be indicative of the smaller change in anisotropy of the potential upon vibrational excitation, compared to Br-HF. This is consistent with the relatively small polarizability of a chlorine atom.

To aid in the assignment of the zero field spectrum to the two isotopes of chlorine, we performed a laser-mass spectrometer experiment designed to determine the origin of the observed spectroscopic signals. This experiment was performed in a second helium nanodroplet spectrometer available in our laboratory, which uses a mass spectrometer in place of the bolometer for monitoring the droplet beam depletion. In this case, vibrational excitation of the complexes upstream of the ionization region of the mass spectrometer again results in a decrease in the corresponding droplet size. This, in turn, reduces the average ionization cross section, providing us with a depletion signal similar to that obtained above using the bolometer. By amplitude modulating the laser and using phase sensitive detection of the mass spectrometer signals, we can obtain mass spectra that are optically labeled, which is to say that the corresponding mass spectra come entirely from droplets containing the species that is being excited by the laser.

This Optically Selected Mass Spectrometry (OSMS) is used in the present study to confirm that the doubling observed in the Cl-HF spectrum is the result of isotopic splitting. Indeed, by positioning the laser to the high frequency side of the Q branch (3887.57cm$^{-1}$), indicated by arrow **B** in Figure 15, we only observe a strong peak in the mass spectrum (Figure 15b) corresponding to the $^{35}$Cl-HF complex, namely at 36 amu (reaction occurs during the ionization of $^{35}$Cl-HF to form H$^{35}$Cl$^+$, discussed below). In contrast, when the laser is tuned to the low frequency side of the spectrum (3887.50 cm$^{-1}$), indicated by arrow **A** in Figure 15, we observe peaks at both 36 and 38 amu (Figure 15a). We see both masses in this case simply because the signal levels (in the optical spectrum) associated with the minor isotope ($^{37}$Cl-HF) are rather low and there is still significant contribution to the spectrum from the wing of the $^{35}$Cl-HF Q branch. These OSMS measurements provide conclusive evidence that the proper assignment of the splitting in the observed spectrum of Cl-HF is to the two isotopes.

A detailed overview of electron impact mass spectrometry in helium droplets has been given in the literature.[88-91] Given the observed ionization reaction however, we briefly discuss its origin. Electron bombardment of the droplet results in the formation of a helium cation that "hops" towards the center of the droplet by resonant charge transfer.[92] Along its way, it may form a (He$_n$)$^+$ cluster (small peaks at mass 32, 36, and 40 in Figure 15) or transfer its charge to Cl-HF. Charge localization on the HF or Cl fragment promotes two reactions (HF$^+$ + Cl → HCl$^+$ + F and Cl$^+$ + HF → HCl$^+$ + F), explaining the appearance of HCl$^+$. Despite the large endothermicity of the second reaction (-1450 cm$^{-1}$ vs. 9700 cm$^{-1}$)[93] the excess energy (dictated by the difference in ionization potentials: IP$_{He}$ - IP$_{HF}$ or IP$_{He}$ - IP$_{Cl}$ ~80000 cm$^{-1}$) released in the charge transfer (from He$^+$ to (Cl-HF)$^+$) may drive either reaction.

Having established that the H-F vibrational origins for the two isotopomers of Cl-HF ($\Delta\nu$ = 0.0380 cm$^{-1}$) are slightly different, it is worth discussing the possible sources of this difference. It is interesting to note that *ab initio* calculations predict the H-F vibrational frequencies for $^{35,37}$Cl-HF, at the equilibrium linear geometry, to have an isotope splitting of less than $1\times10^{-4}$ cm$^{-1}$. Although these calculations cannot be trusted to this absolute accuracy, they do indicate that the direct mass effect is much too small to account for the observations. Instead, we turn our attention to the fact that the complexes are not rigidly linear, but undergo significant intermolecular bending. This we already know from the dipole moment measurements presented above. Indeed, the bending potential for these systems are presumed to be rather broad, consistent with that calculated for Cl-HCl,[23] meaning that the complex undergoes significant intermolecular bending. If the effective mass for this bending motion is different for the two isotopomers, the corresponding difference in the bending amplitude will change the associated intramolecular vibrational averaging, resulting in slightly different vibrational origins for the two isotopomers.

Flygare and co-workers[94] have discussed this issue previously for closed shell complexes, including the $^{82,83,84,86}$Kr-HF isotopomers. From the microwave spectrum they estimate the vibrationally averaged bending angle (VAA) from the projection of the nuclear spin-spin interaction constant in HF onto the A inertial axis of the complex. For $^{82}$Kr-HF and $^{86}$Kr-HF the VAA's were estimated to be 38.67° and 37.86° respectively. Infrared measurements[95] on the HF stretch of this complex show an isotope splitting on the order of 0.005 cm$^{-1}$, which is attributed to this effect.

Given the magnitude of the isotope splitting in the chlorine spectra ($\Delta\nu$ = 0.0380 cm$^{-1}$), we do not expect a significant isotope splitting for Br-HF, due to the much smaller percent mass change in this case. Indeed, the ratios of the reduced masses in the two systems differ by a factor of four. This fact alone (assuming the bending potentials are the same) would make the isotope effect on the vibrational frequency a factor of $\sqrt{4}$ smaller for Br-HF (0.019 cm$^{-1}$). Within experimental resolution, a splitting of this magnitude is not evident in the field free spectrum. However, a careful examination of several pendular scans of Br-HF does lend supporting evidence to an isotope splitting of approximately this value. Figure 16a shows a spectrum of Br-HF recorded with an applied electric field strength of 43.4 kV/cm. The simulated spectra corresponding to only one isotope (labeled C) has quite poor agreement of the intensities for the unresolved bands. A sum of two such bands separated by 0.010cm$^{-1}$ is shown (B) and the agreement is found to be excellent, lending support for an isotope splitting of this magnitude. In the field free spectrum, this splitting only contributes to a slightly larger observed linewidth.

**Summary**

We reported here the direct observation of heavy-light-heavy complexes consisting of halogen atoms (Cl, Br, and I) with a single HF molecule, formed in helium nanodroplets. Infrared spectroscopy is used to determine the structures of these complexes to be (vibrationally averaged) linear, in agreement with theoretical calculations. Fine structure is observed in the spectra corresponding to isotope splittings for Cl-HF and Br-HF (Iodine has only one naturally occurring isotope) and nuclear magnetic hyperfine interactions. The isotope splittings are in general agreement with the closed shell analogues Kr-HF and Xe-HF, and as such are sensitive to the percent change in mass for each case. The splitting between these two bands provides a stringent test for theoretical surfaces and we hope this comparison is made once fully-dimensional high-quality theoretical potential energy surfaces become available. The observed nuclear magnetic hyperfine interactions are also in general agreement with that found in the isolated atoms. Dipole moments of these complexes have been measured using Stark spectroscopy and are considerably larger than that recently predicted by bound state calculations, showing that the induced dipoles of these polarizable halogen atoms cannot be neglected for quantitative accuracy. It would be interesting to include these effects in the theoretical calculations so that better comparisons could be made between experiment and theory.

Concurrent with these studies was the pursuit of F-HF. No evidence for its formation has been observed to date, possibly calling into question the assignment of the argon matrix work.[31] It may be that the growth process in helium favors different geometrical structures, none of which were observed in this study. Given the importance of understanding the complex potentials of these systems, and the now proven ability of helium droplets to stabilize species of this type, such systems will be an active area of future work.

It has been proposed[28] that complexes of this type may posses a T-shaped structure, but display several of the properties of a linear open-shell molecule. Given the very large frequency shift from the HF monomer however, this is unlikely the case for the I-HF, Br-HF and Cl-HF complexes observed here. The dynamics of these systems can certainly be regarded as floppy, as our results have shown for the disagreement in the dipole moments and vibrational averaging, however we feel the linear equilibrium view is still a good one. The excellent agreement with both the nuclear hyperfine structure and the Stark effect also support our assignment. These theories have only included the effects believed to be present in a linear molecule case, giving very good results. In addition, the Stark effect is extremely sensitive to the relative orientation of the permanent dipole moment and the vibrational transition moment. If the two moments were not parallel (as in the linear molecule case) the resulting Stark spectrum would be expected to have quite different character. Nevertheless, the fact that this point is still the subject of considerable debate illustrates how poorly we still understand these species, calling out for more experiments and theory on these entrance channel complexes.

**Acknowledgments**

The authors wish to thank Dr. Markus Meuwly and Dr. Jeremy Hutson for their theoretical work on Br-HF[60] and its correspondence before publication. JMM also wishes to thank Dr. John Brown for helpful discussions concerning the inclusion of the Stark effect in an open shell linear rotor program. This work was supported by the Air Force Office of Scientific Research (AFOSR). Partial support is also acknowledged from the National Science Foundation (CHE-99-87740) and the Alexander von Humboldt Foundation (fellowship for J.K.).

Table 1) A summary of optimized stationary points for the hydrogen bonded (global minimum) configuration of X-HF. Structural parameters, along with other molecular constants, were calculated at the UMP2 and UCCSD(T) level of theory with the Aug-cc-pVTZ basis set in Gaussian 03. Binding energies have been corrected for basis set superposition error using the method of Boys and Bernardi.[63] Dipole moments have been calculated using the UMP2 electron density. Calculations labeled by (ECP) were performed at the UMP2 level using a pseudopotential basis set for bromine and iodine, see text for details. Scaled frequencies were obtained by comparing a free HF calculation with the vibrational frequency observed in helium droplets. The argon matrix data comes from the work of Andrews.[15]

| UMP2 (UCCSD(T)) Aug-cc-pVTZ | HF | F-HF | Cl-HF | Br-HF | Br-HF (ECP) | I-HF (ECP) |
|---|---|---|---|---|---|---|
| $D_e$ (cm$^{-1}$) | — | 364 (360) | 672 (676) | 721 (720) | 645 | 700 |
| $\nu_{harmonic}$ (cm$^{-1}$) | 4123 (4124) | 4096 (4103) | 4030 (4045) | 3998 (4017) | 3985 | 3973 |
| $\nu_{scaled}$ (cm$^{-1}$) | 3959.19 | 3933 (3939) | 3870 (3883) | 3839 (3856) | — | — |
| $\nu_{Ar}$ (cm$^{-1}$) | 3962 | 3908 | 3858 | 3831 | — | — |
| $R_{X-H}$ (Å) | — | 2.087 (2.068) | 2.371 (2.378) | 2.439 (2.449) | 2.392 | 2.611 |
| $R_{HF}$ (Å) | 0.9218 (0.9210) | 0.9232 (0.9222) | 0.9258 (0.9246) | 0.9272 (0.9258) | 0.9278 | 0.9283 |
| $B$ (cm$^{-1}$) | 20.70 (20.73) | 0.195 (0.198) | 0.125 (0.124) | 0.0953 (0.0948) | 0.098 | 0.080 |
| $\langle\mu\rangle_{eq}$ (D) | 1.81 | 2.05 | 2.34 | 2.48 | 2.52 | 2.63 |

Table 2) Optimized structures of the linear HF-X complexes. The dissociation energies listed have not been corrected for basis set superposition error. Calculation of this correction results in the species being unbound however. This result calls into question the validity of the calculation and the existence of a potential minimum for this geometry. No imaginary frequencies were calculated.

| UMP2 Aug-cc-pVTZ | HF-F | HF-Cl | HF-Br |
|---|---|---|---|
| $D_e$ (cm$^{-1}$) | 6 | 20 | 70 |
| $\nu_{harmonic}$ (cm$^{-1}$) | 4123 | 4121 | 4120 |
| $R_{X-F}$ (Å) | 3.416 | 3.682 | 3.718 |
| $R_{HF}$ (Å) | 0.9218 | 0.9219 | 0.9220 |
| B" (cm$^{-1}$) | 0.143 | 0.095 | 0.074 |
| $\langle\mu\rangle_{eq}$ (D) | 1.83 | 1.90 | 1.94 |

Table 3) A summary of optimized structures for the "T-shaped" X-HF complexes calculated at the UMP2/Aug-cc-pVDZ level. The angle theta is defined by using the F atom (in HF) as the vertex. Binding energies have been corrected for basis set superposition error.

| UMP2 Aug-cc-pVTZ | F-HF | Cl-HF | Br-HF |
|---|---|---|---|
| $D_e$ (cm$^{-1}$) | 220 | 375 | 404 |
| $\nu_{harmonic}$ (cm$^{-1}$) | 4116 | 4109 | 4107 |
| $R_{X-F}$ (Å) | 2.662 | 2.970 | 3.036 |
| $R_{HF}$ (Å) | 0.9223 | 0.9228 | 0.9229 |
| A", B", C" (cm$^{-1}$) | 23.7, 0.241, 0.238 | 24.5, 0.148, 0.147 | 26.0, 0.113, 0.112 |
| $\langle\mu\rangle_{eq}$ (D) | 1.83 | 1.87 | 1.92 |
| Theta (deg) | 109.8 | 112.4 | 116.2 |

Table 4) Tabulated results for the two isomers of Br$_2$-HF. Two distinct isomers are observed in helium droplets in agreement with *ab initio* (MP2/Aug-cc-pVDZ) calculations (See Figures 2 and 3). Br$_2$-HF represents the "L" shaped isomer, while Br$_2$-FH is the "bent" one. Calculated frequencies have been scaled to the HF monomer. The experimental constants are those obtained from fitting the corresponding spectra and are compared with the *ab initio* values.

| Constant (cm$^{-1}$) | Helium droplet | MP2 Aug-cc-pVDZ |
|---|---|---|
| Br$_2$-FH | | |
| $\nu_0$ | 3941.47 | 3940.98 |
| (B+C)/2 | 0.009 | 0.032 |
| $D_J$ | 1.0×10$^{-6}$ | — |
| Br$_2$-HF | | |
| $\nu_0$ | 3874.14 | 3859.16 |
| A', (B'+C')/2 | 0.018, 0.01 | — |
| A", (B"+C")/2 | 0.020, 0.01 | 0.108, 0.048 |
| $D_J$ | 1.0×10$^{-6}$ | — |

Table 5) A summary of the experimental parameters for the linear X-HF complexes, as obtained from fitting the infrared spectra. Isotope splittings of 0.038 cm$^{-1}$ and 0.010 cm$^{-1}$ were observed for the Cl-HF and Br-HF complexes, respectively. In each case the heavier isotope is shifted to the red due to vibrational averaging (see text).

| Constant | $^{35}$Cl-HF | Br-HF | I-HF |
|---|---|---|---|
| $\nu_0$ (cm$^{-1}$) | 3887.54 | 3867.46 | 3847.82 |
| B (cm$^{-1}$) | 0.055 | 0.043 | 0.037 |
| D (cm$^{-1}$) | $1.2 \times 10^{-4}$ | $0.95 \times 10^{-4}$ | $0.3 \times 10^{-4}$ |
| [a$\Lambda$ + (b+c)$\Sigma$] (cm$^{-1}$) | 0.005 | 0.045 | 0.035 |
| $\mu$ (D) | 1.9 | 2.1 | 2.2 |
| $\Delta\mu$ (D) | 0 | +0.25 | +0.2 |

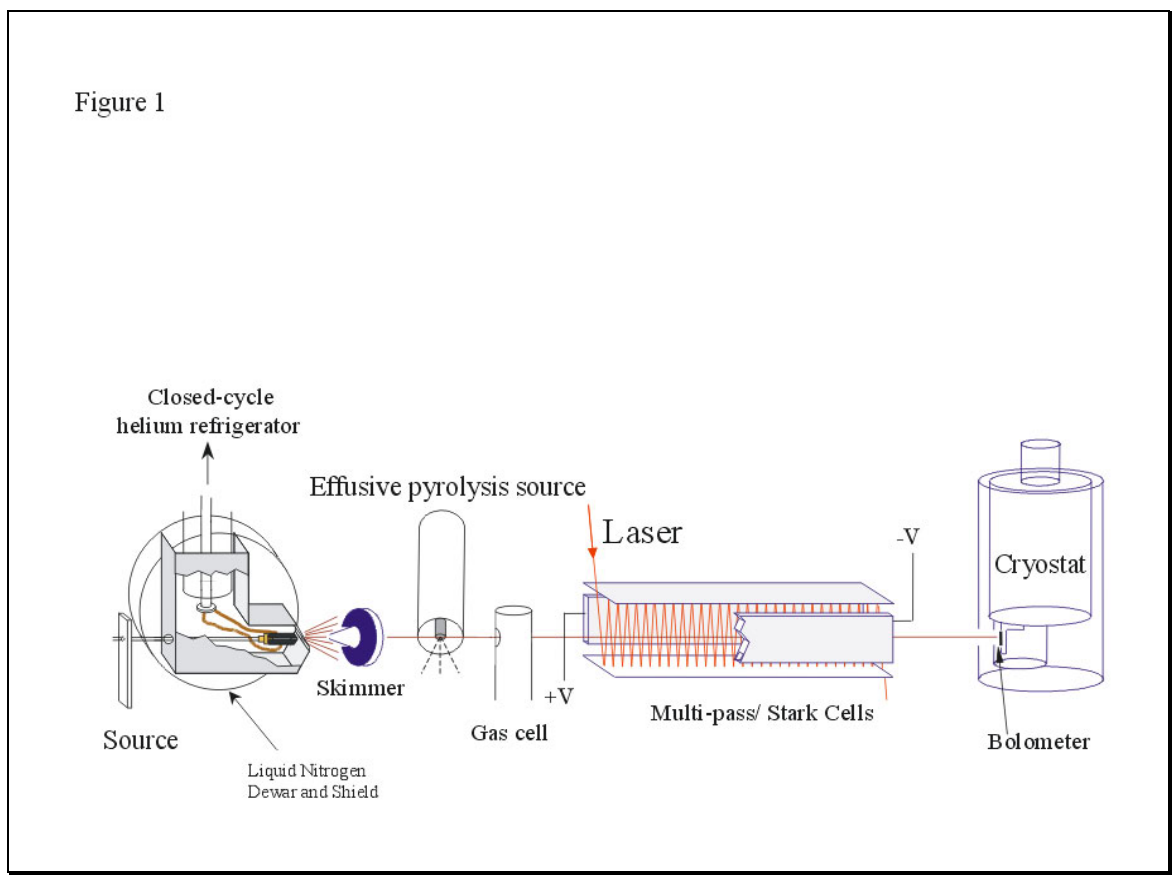

Fig 1) Schematic diagram of the experimental apparatus used in the present study. A bolometer detector was used to monitor the laser-induced depletion of the helium droplet beam intensity.

Figure 2

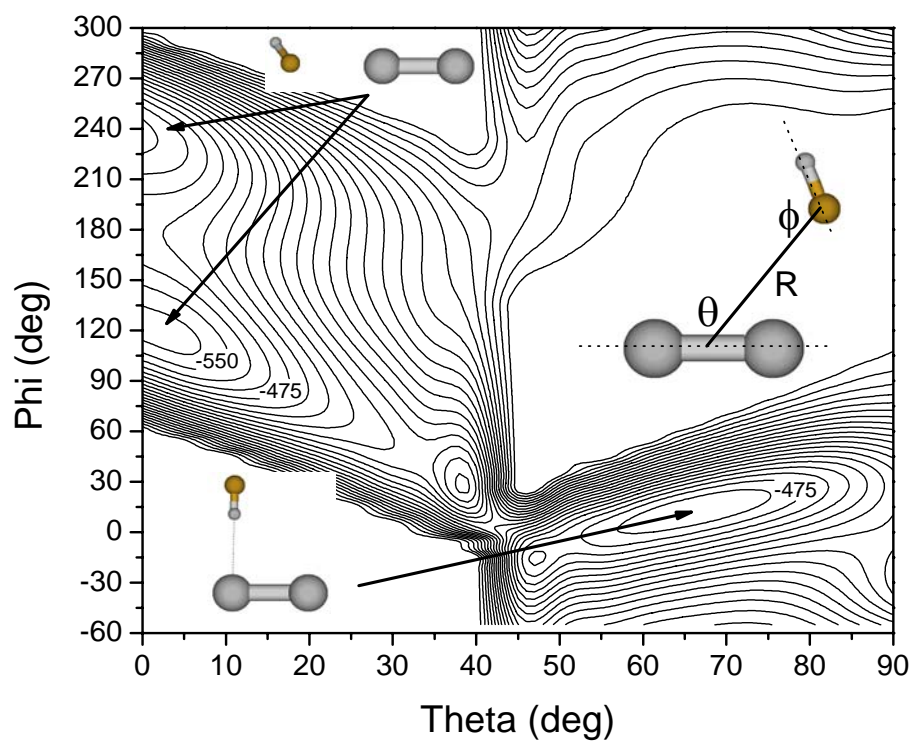

Fig 2) A two-dimensional cut through the MP2/aug-cc-pVDZ PES for the $Br_2$-HF system. At each point, the geometry was fully relaxed (except the two rotor angles) and the binding energy corrected for basis set superposition error by counterpoise correction. The surface has two minima that correspond to the "bent" and "L" shaped isomers drawn as insets. The corresponding isomers are observed in the helium droplet experiments reported here. Neighboring contours are spaced by 25 $cm^{-1}$.

Figure 3

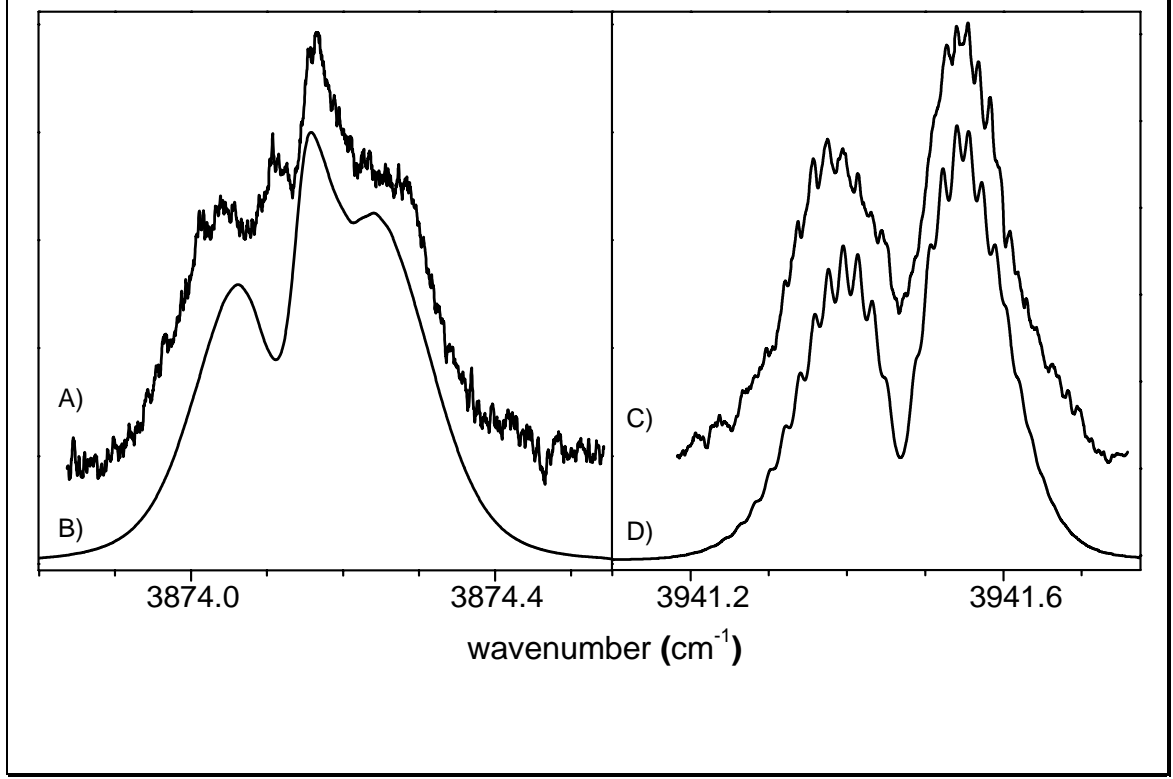

Fig 3) Zero field spectra, (A) and (C), of the L-shaped and bent isomers of $Br_2$-HF, respectively. Inertial parameters derived from the simulation of the spectra, (B) and (D), are shown in Table 4 and are in good agreement with *ab initio* calculations. A rotational temperature of 0.37 K was used in the simulations.

Figure 4

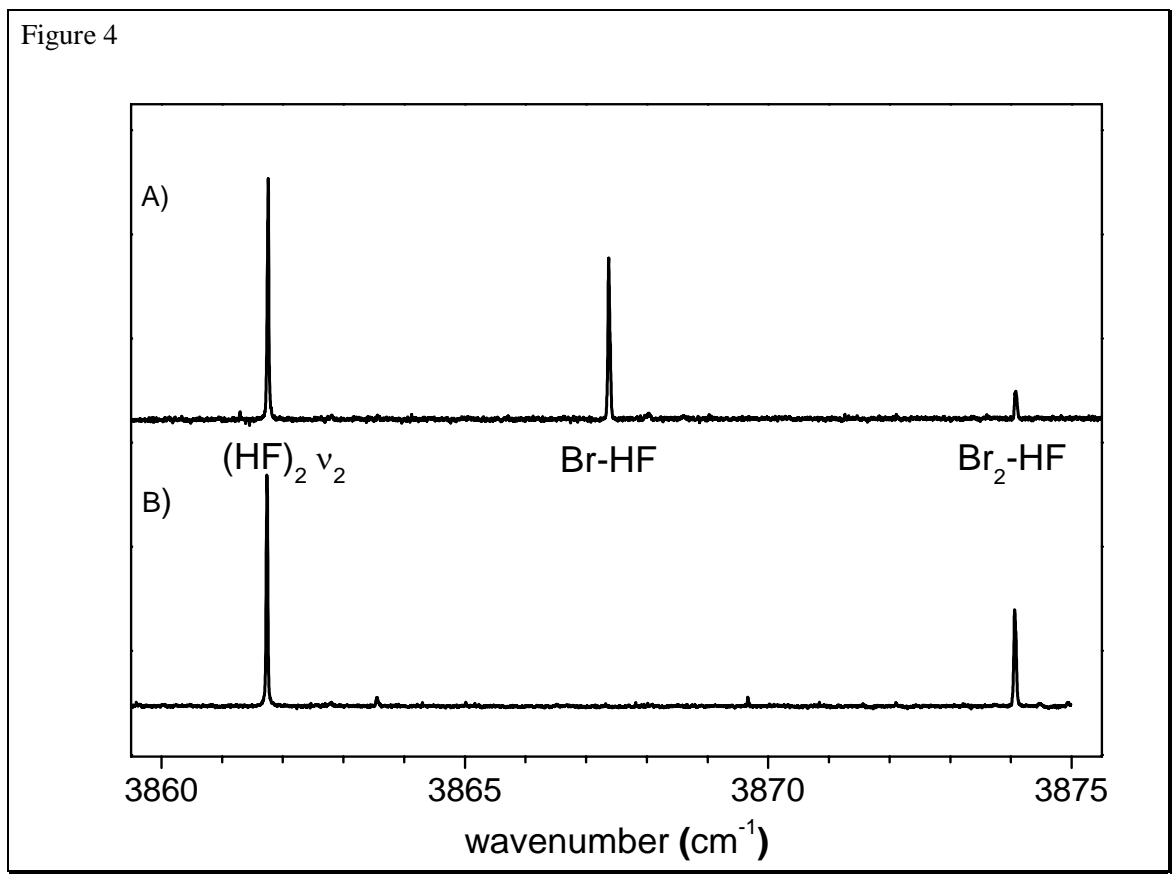

Fig 4) Pendular spectra recorded in the $\nu_2$ HF dimer region, showing the appearance of bromine related complexes. In the bottom trace the pyrolysis source is at room temperature and the peak at 3874 cm$^{-1}$ corresponds to the L-shaped isomer of Br$_2$-HF. The peak at 3862 cm$^{-1}$ is due to the HF dimer. The top trace was recorded with the pyrolysis source at 1200 K, were a new peak appears that is assigned to Br-HF. Note that the intensity of the Br$_2$-HF peak is diminished, showing the extent of pyrolysis.

Figure 5

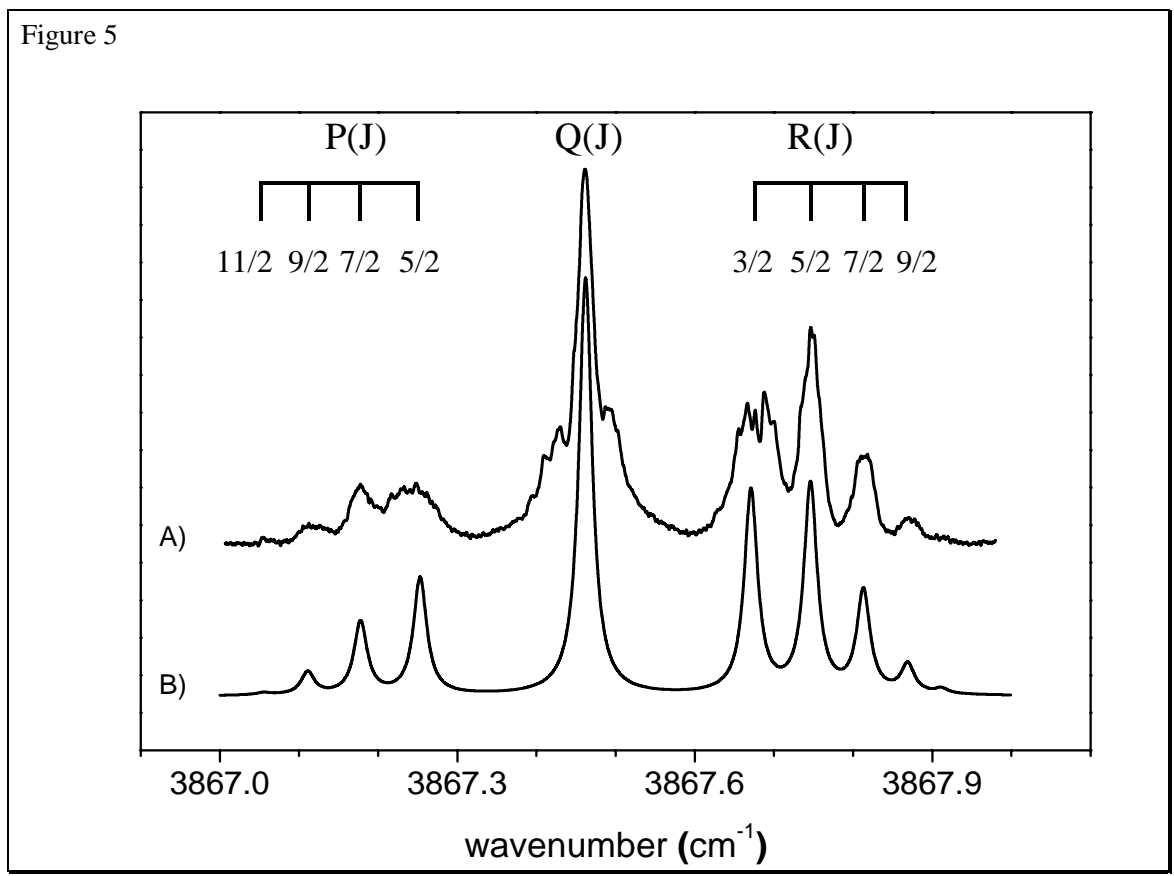

Fig 5) A rotationally resolved spectrum (A) of Br-HF solvated in helium droplets. The spectral shape is consistent with that of a $^2\Pi_{3/2}$ electronic state. Also shown is a simulation (B), based upon the following constants: B' = B" = 0.043 cm$^{-1}$, D' = D" = 0.95× 10$^{-4}$ cm$^{-1}$ and T = 0.35 K.

Figure 6

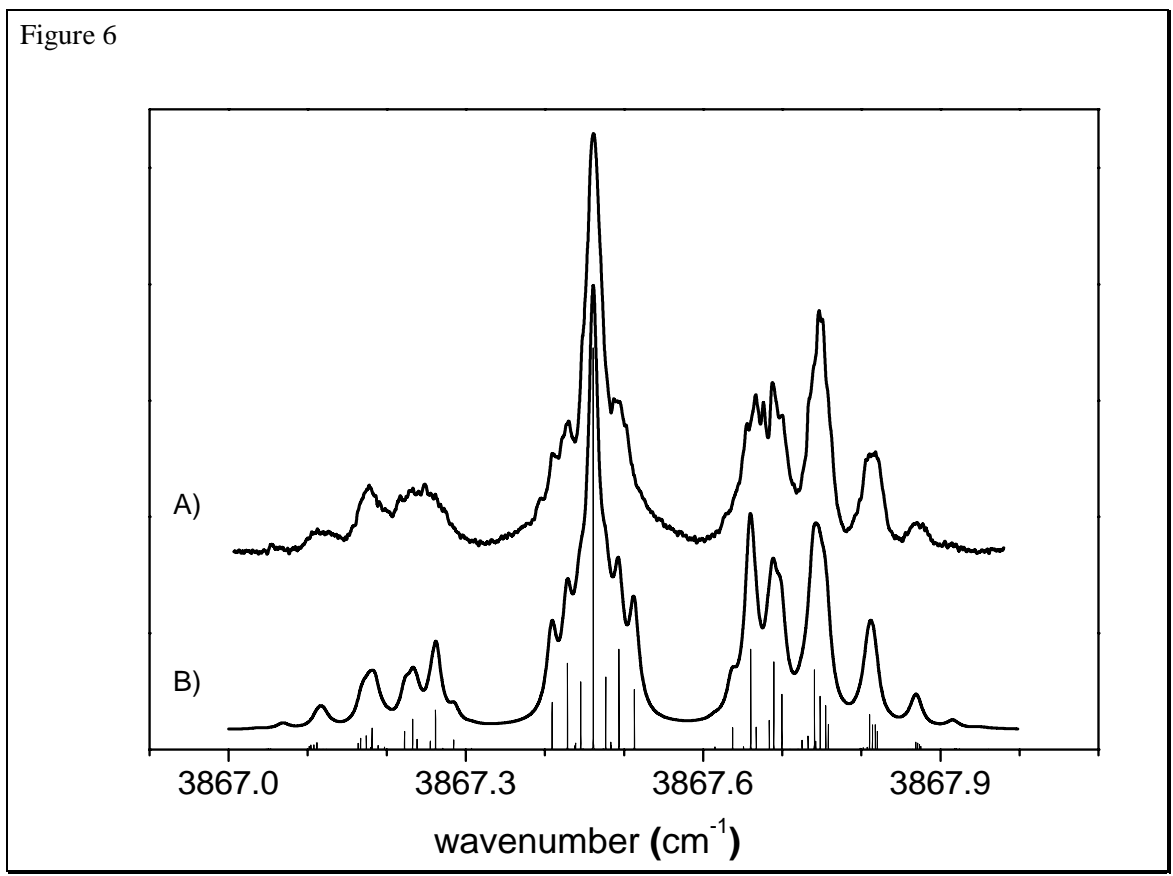

Fig 6) The infrared spectrum of Br-HF (A) shown with a simulation (B) that includes nuclear magnetic hyperfine interactions, I = $^3/_2$, (see text). The stick spectrum shows all of the transitions that lie under the observed features.

Figure 7

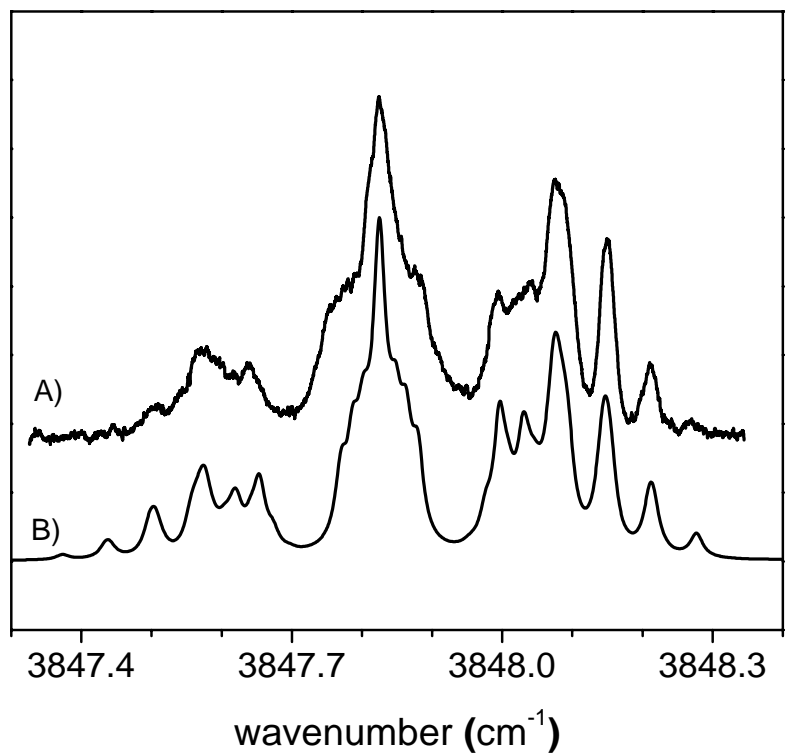

Fig 7) The infrared spectrum of I-HF in helium nanodroplets (A). The simulation (B) includes the effects of nuclear magnetic hyperfine structure, $I = {}^5/_2$ (see text).

Figure 8

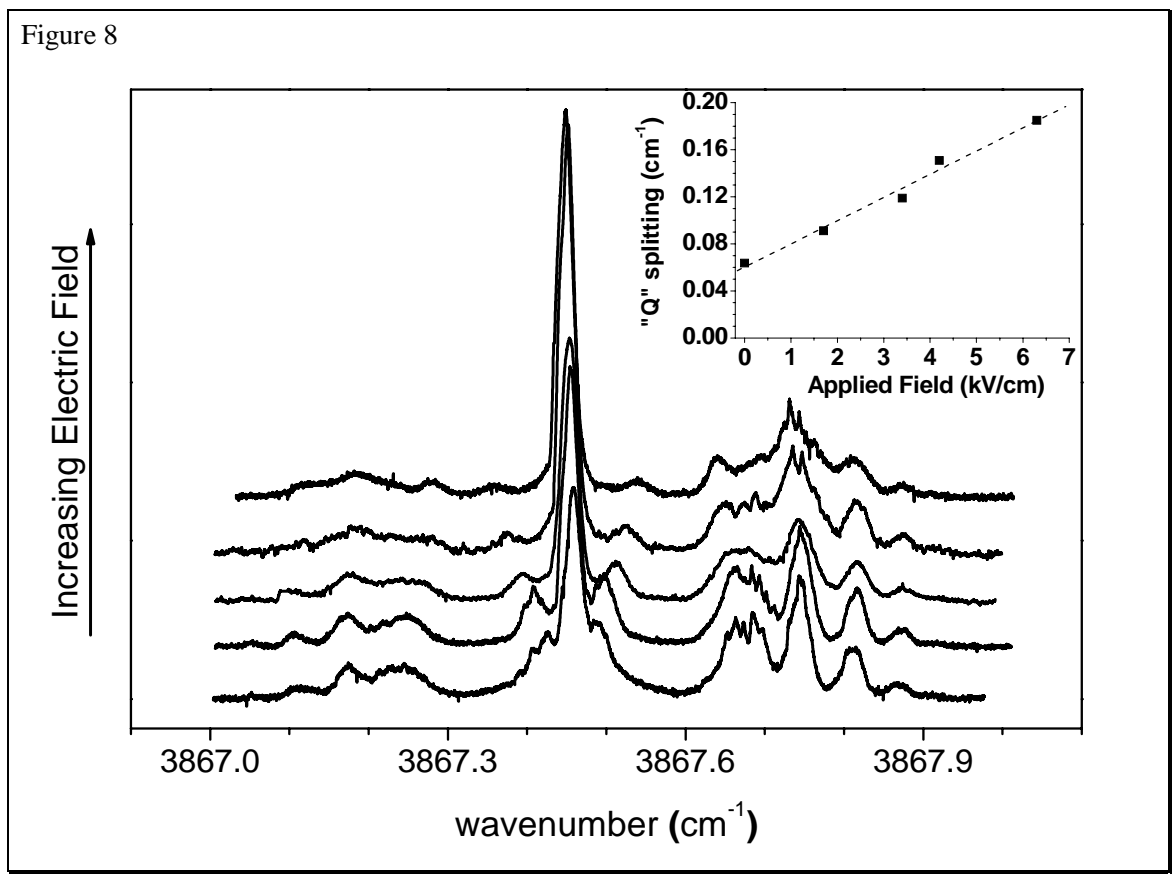

Fig 8) A series of spectra of Br-HF showing the dependence on the magnitude of the applied electric field, in this case aligned parallel to the laser polarization direction. Note that the transitions representing the shoulder on the Q branch split approximately linearly with the field strength, as shown in the inset.

Figure 9

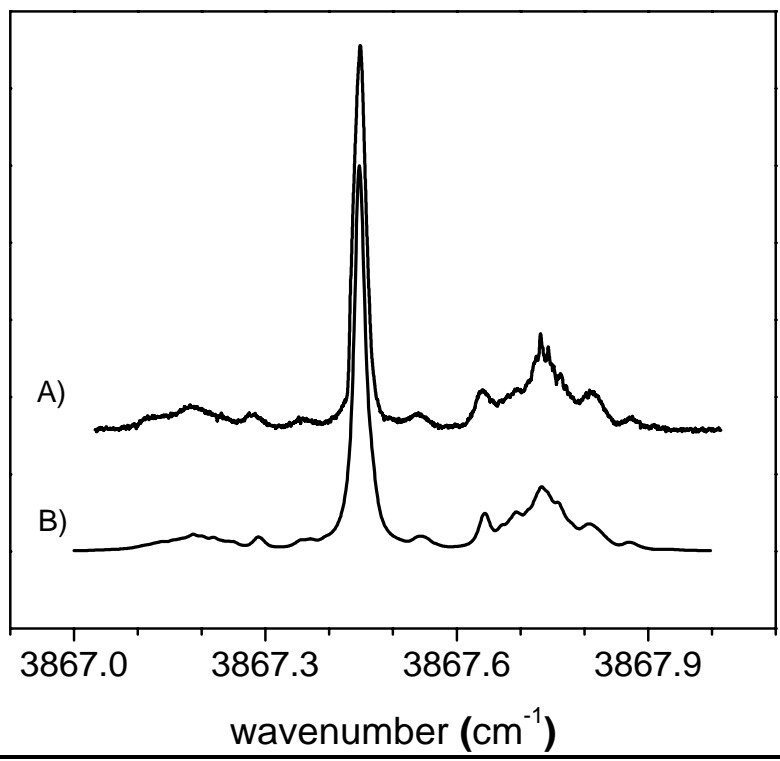

Fig 9) A Stark spectrum (A) of Br-HF, recorded with a 5.17 kV/cm electric field applied parallel to the laser polarization direction. The simulation (B) is based upon a fitted dipole moment of 2.10 D, with an increase in the dipole moment upon vibrational excitation of 0.25 D.

Figure 10

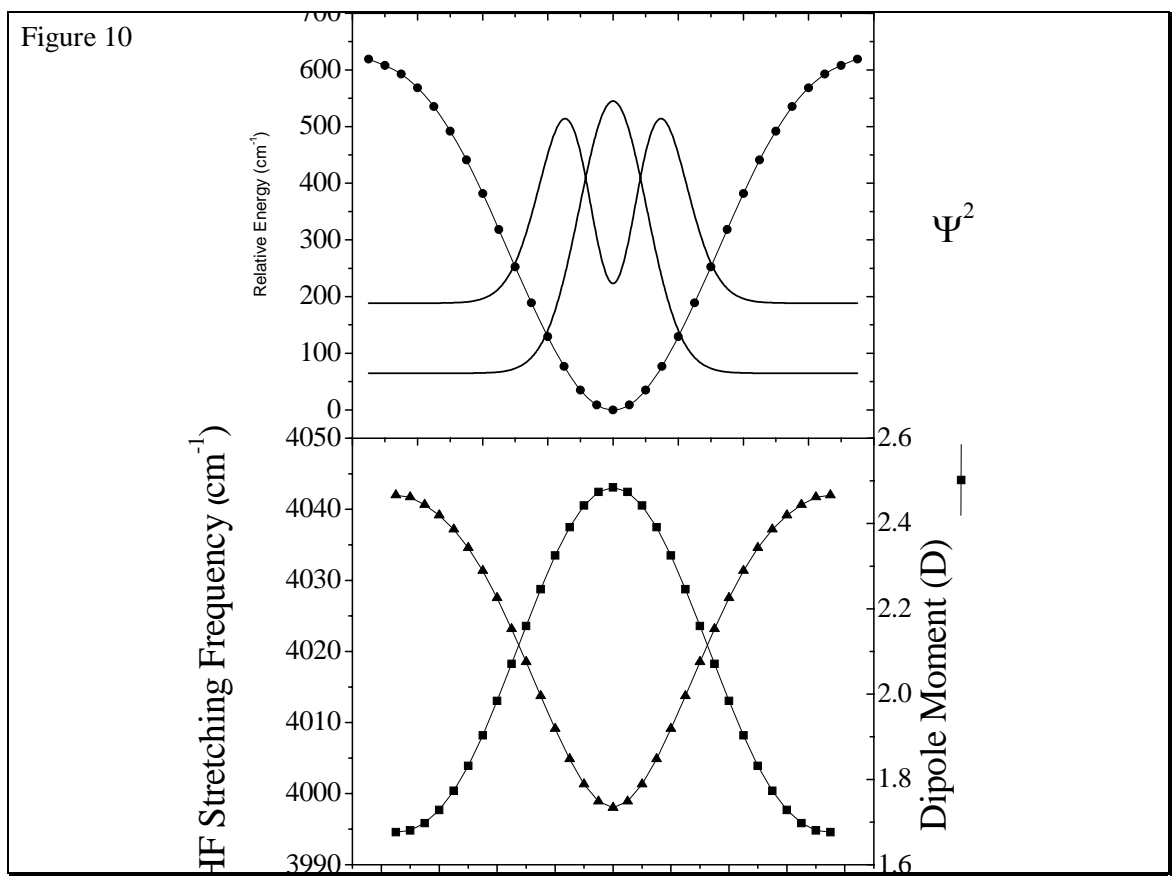

Fig 10) Upper panel: A one-dimensional slice through the *ab initio* (UMP2/aug-cc-pVTZ) potential ($^2A'$) for the Br-HF complex, along the H-F bending coordinate. The ground and first excited state wavefunctions were calculated (see text), and $\Psi^2$ is shown for reference. Lower panel: *Ab initio* calculations of the HF stretching frequency (Triangles) and the permanent dipole moment (Squares) as a function of the bending angle. The effects of vibrational averaging are found to lower the dipole moment and shift the HF stretching frequency to the blue, in agreement with experimental results.

Figure 11

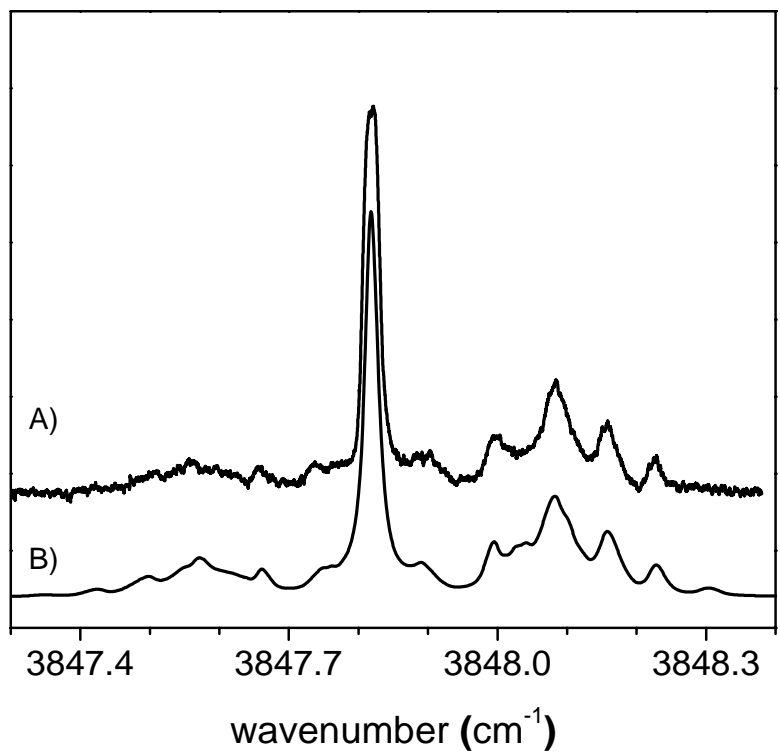

Fig 11) An experimental Stark spectrum of I-HF (A) recorded at an electric field strength of 3.86 kV/cm. The spectrum (B) was fit to the experimental data, yielding a dipole moment of 2.2 D. The increase in dipole moment upon vibrational excitation is 0.2 D.

Figure 12

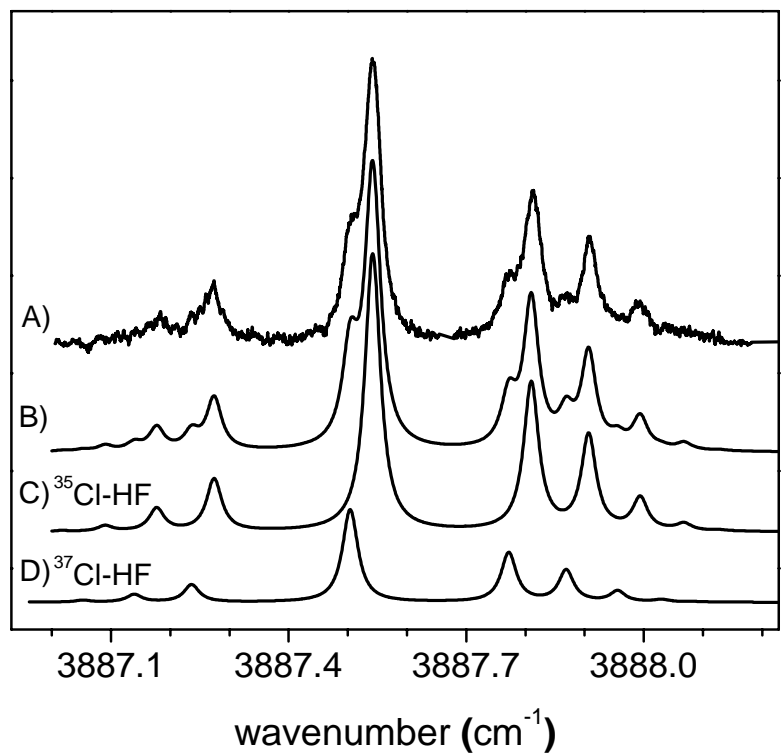

Fig 12) The experimental infrared spectrum (A) of Cl-HF in helium nanodroplets. The observed splitting is attributed to the two naturally occurring isotopes of chlorine. A good fit (B) is obtained using spectra (C) and (D), separated by 0.0380 cm$^{-1}$. The relative intensities of the two bands are scaled to the natural abundances of the two isotopes of chlorine. In addition, a nonzero nuclear magnetic hyperfine constant was necessary to reproduce the measured rotational line intensities. The simulation parameters are given in Table 5.

Figure 13

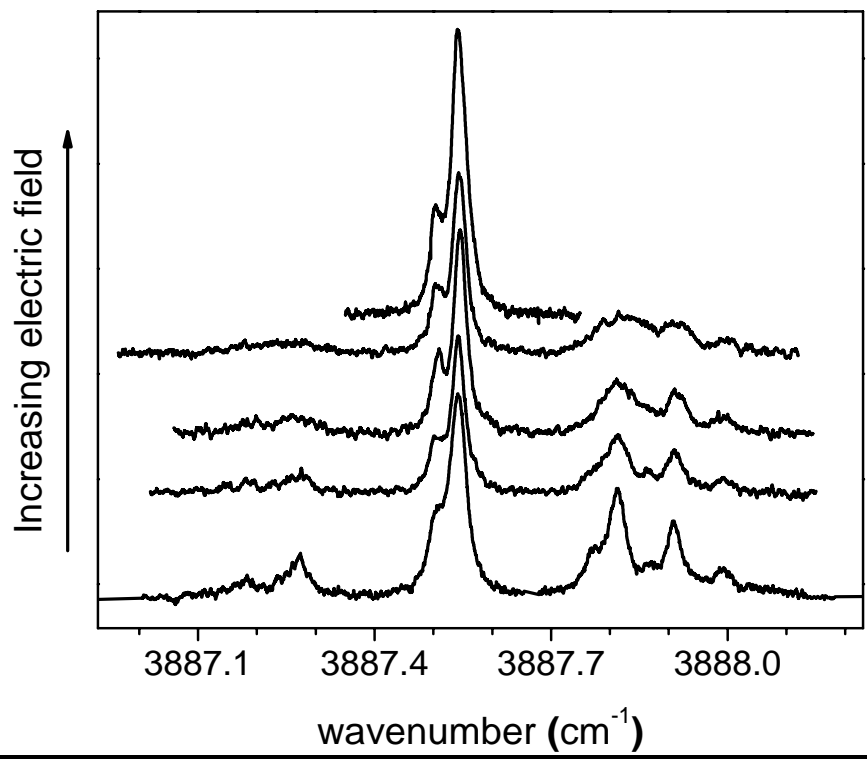

Fig 13) A series of spectra showing the evolution of the Cl-HF spectra as a function of applied electric field strength.

Figure 14

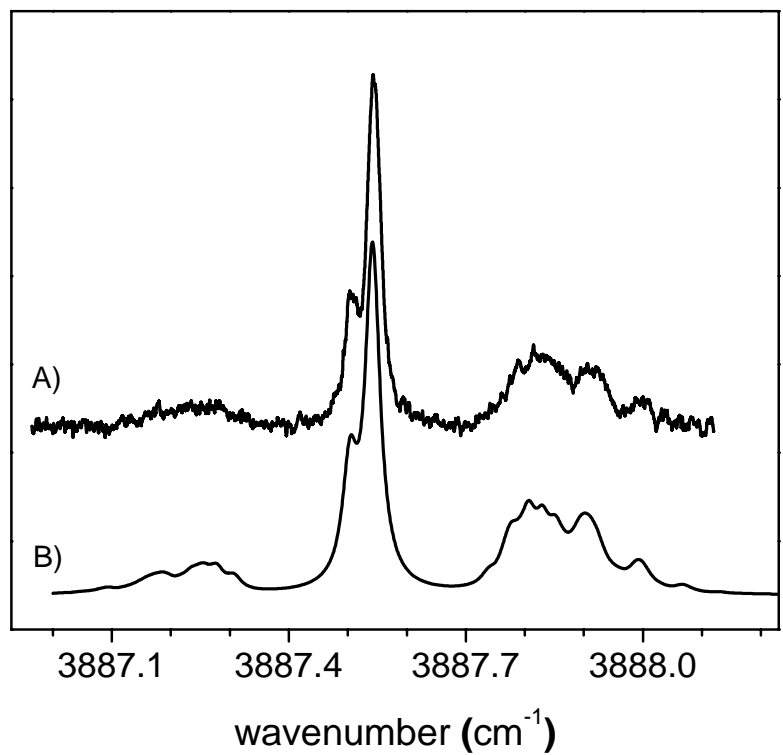

Fig 14) A Stark spectrum (A) of Cl-HF recorded at an electric field strength of 3.51 kV/cm applied parallel to the laser polarization direction. The fitted spectrum (B) yielded an experimental dipole moment of 1.9 D. The simulation includes contributions from both isotopes of Chlorine.

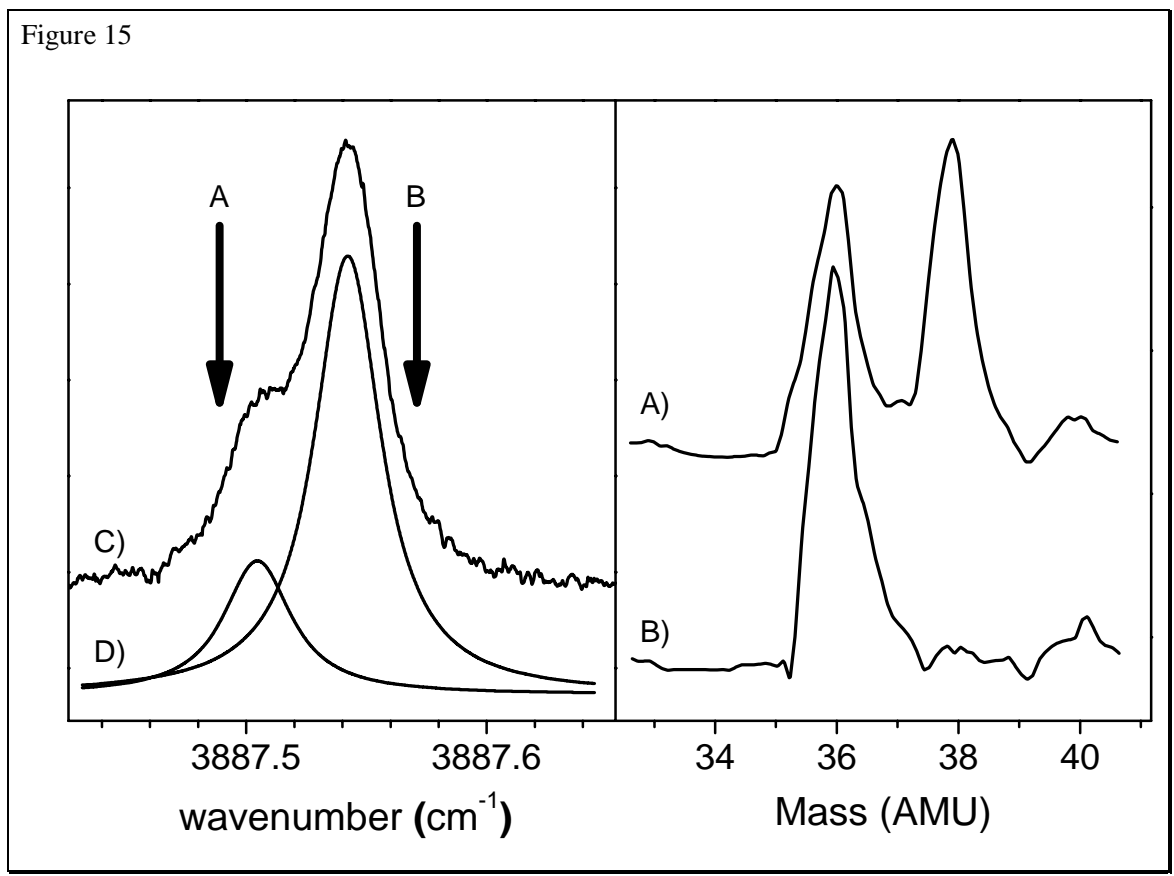

Fig 15) Left panel: An expanded view of the Q branch region of the Cl-HF spectrum shown in Figure 12. The vertical arrows show the laser frequencies used to record the two mass spectra shown in the right panel. The two Lorentzian line shapes shown in (D) indicate the relative contributions from the $^{35}$Cl and $^{37}$Cl isotopomers at these laser frequencies. Right panel: Optically Selected Mass Spectra (OSMS) of Cl-HF in helium droplets. This technique is used to confirm that the observed splitting in the Cl-HF infrared spectrum is due to the two isotopes of chlorine (see text).

Figure 16

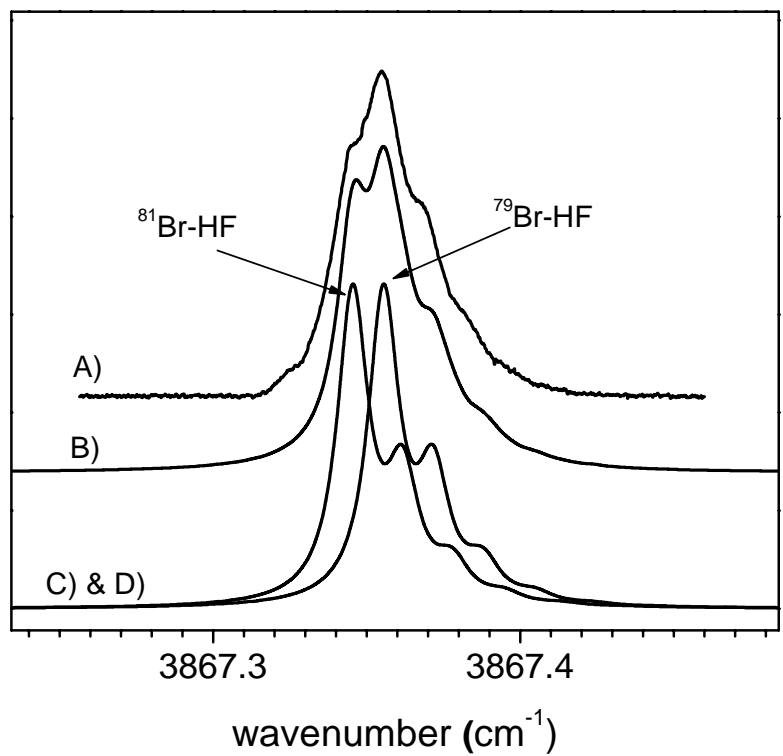

Fig 16) A pendular spectrum of Br-HF recorded at an applied electric field strength of 43.4 kV/cm. The fine structure in this band is well described (B) by an isotope splitting of 0.01 cm$^{-1}$. The smaller isotope splitting, in comparison with Cl-HF, is consistent with the correspondingly smaller ratio of the reduced masses for Br-HF. The two simulated spectra, (C) and (D), were generated using the constants derived from fitting the field free and Stark spectra.